# A Comparative Analysis of Progressive Multiple Sequence Alignment Approaches using UPGMA and Neighbor Join based Guide Trees

Dega Ravi Kumar Yadav[1] and Gunes Ercal[2]

Department of Computer Science, SIUE, Edwardsville, Illinois, USA

*ABSTRACT*

*Multiple sequence alignment is increasingly important to bioinformatics, with several applications ranging from phylogenetic analyses to domain identification. There are several ways to perform multiple sequence alignment, an important way of which is the progressive alignment approach studied in this work. Progressive alignment involves three steps: find the distance between each pair of sequences; construct a guide tree based on the distance matrix; finally based on the guide tree align sequences using the concept of aligned profiles. Our contribution is in comparing two main methods of guide tree construction in terms of both efficiency and accuracy of the overall alignment: UPGMA and Neighbor Join methods. Our experimental results indicate that the Neighbor Join method is both more efficient in terms of performance and more accurate in terms of overall cost minimization.*

## 1. INTRODUCTION

The traditional pairwise sequence alignment problem in its utmost generality is to find an arrangement of two given strings, S and T, such that the arrangement yields information on the relationship between S and T, such as the minimum number of changes to S that would transform S into T. In the context of DNA sequences, which can be viewed as strings from the 4 letter alphabet {A, C, G, T}, these changes may represent mutation events, so that the alignment sought yields important evolutionary information [15]. Similarly, the pairwise sequence alignment problem can be generalized to the *multiple* sequence alignment problem to yield information on the relatedness of multiple sequences. Applications of the multiple sequence alignment (MSA) problem for DNA sequences include phylogenetic analysis, domain identification, discovery of DNA regulatory elements, and pattern identification. Additionally, MSA applications for protein sequences also includes protein family identification and structure prediction. This work is concerned with approaches to multiple sequence alignment in the context of DNA sequences.

Generally, aligning two sequences is straightforward via dynamic programming. But pairwise alignment is insufficient for many applications in which the relationship among several sequences is sought. Moreover, it is infeasible to naturally extend the dynamic programming approach that works for pairwise sequence alignment directly to multiple sequence alignment when there are more than three sequences to align. Unfortunately, multiple sequences alignment is NP-hard based on SP (sum-of-pairs) scores [1]. Therefore, heuristics are crucial to MSA.

Progressive alignments are by far the most widely used heuristic multiple sequence alignment method [2, 3]. Progressive alignment is done in three major steps; 1. Perform pairwise distance calculations for all the input DNA sequences. 2. Build the guide tree using the distance matrix computed in the previous step. 3. Based on the guide tree, perform progressive alignment with nearest sequences first. In this project, we compare two important progressive alignment approaches to MSA in terms of algorithmic efficiency as well as alignment accuracy, namely progressive MSA with UPGMA based guide trees and progressive MSA with Neighbor Join based guide trees. Our results indicate that the Neighbor Join method of guide tree construction is preferable in terms of both efficiency and accuracy of the overall resulting MSA.

## 2. BACKGROUND AND RELATED WORK

The main steps of the progressive alignment methodology are as follows [6, 13]:
1. Compute pairwise distances for all the sequences.
2. Build the guide tree based on the distance matrix.
3. Align first two sequences based on the guide tree leaf nodes using dynamic programming (Global Alignment).
4. From the next sequence alignment, construct a *profile* and align the new sequence to the profile.
5. Repeat step 4 until the longest branch leaf node is aligned. Hence, MSA is achieved.
6.

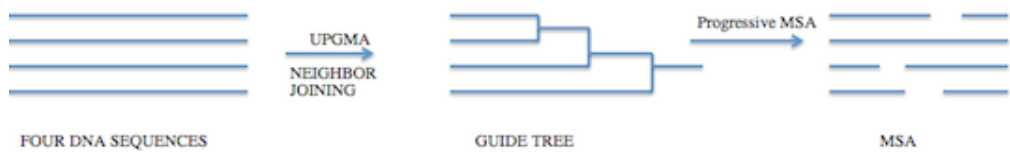

Figure 1: Progressive Multiple sequence Alignment

### 2.1. Pairwise Distance

The distance between two DNA sequences is known as the pairwise distance. In this project, a distance matrix is computed among the species using Jukes Cantor distance formula. There are a number of evolutionary models proposed to measure pairwise distance, the first of which is the Jukes-Cantor model [8]. The Jukes Cantor distance formula is based on the ratio of number of matches to the number of non-gaps in the DNA of two sequences of the species. Jukes cantor distance formula is:

$$D = (-3/4 * \log (1-(4/3*R)))$$

Here `D' is the Jukes-Cantor distance between two DNA sequences and `R' is the ratio of number of matches to the number of non-gap letters [8]. Once we obtain a distance matrix for the specified species, a guide tree based on distance matrix is built.

### 2.2. Guide Tree

In progressive alignment, the guide tree plays an important role as the branches of the tree determine which sequence is to be considered for the next step of alignment [10]. A guide tree is a phylogenetic tree that is constructed dependent on the distance matrix of the DNA sequences. A phylogenetic tree is an evolutionary tree which shows interrelations among various biological species. Phylogenetic trees are dependent on physical or genetic characteristics similarities and differences. They show the distance between pairs of sequences when the tree edges are weighted [10]. There are several types of phylogenetic trees; rooted, unrooted, and bifurcating. In this project we will be using an unrooted phylogenetic tree as our guide tree. Guide trees may be built using clustering algorithms or other learning models. In this project, the guide tree is constructed using both UPGMA and Neighbor-Joining algorithms and the final results are compared. Now we will have a closer look at these two algorithms.

### 2.2.1. UPGMA

UPGMA stands for Un-weighted Pair-Group Method with Arithmetic mean. Un-weighted refers to all pairwise distances contributing equally, pair-group refers to groups being combined in pairs, and arithmetic mean refers to pairwise distances between groups being mean distances between all members of the two groups considered [7].

Consider four DNA sequences namely: $S_1$, $S_2$, $S_3$ and $S_4$. First, find the pairwise distances between all the sequences. Then, find the smallest value in the distance matrix and its corresponding sequences of the shorter distance. For instance let the two sequences with the shortest distance between them be $S_1$ and $S_2$. Now, cluster $S_1$ and $S_2$ and name the cluster as $C_1$, updating the distance matrix by eliminating $S_1$ and $S_2$, but including $C_1$. The $C_1$ value corresponding to the remaining sequences in the distance matrix is calculated with the values of $S_1$ and $S_2$, i.e., arithmetic mean of $S_1$ and $S_2$ distances with corresponding to the other sequences. Now, moving forward by considering the updated distance matrix, find the smallest distance again and its corresponding sequences or clusters (namely sets of sequences). Say this next smallest distance corresponds to that between clusters $C_3$ and $C_4$. Then, in the next step, $C_3$ and $C_4$ would be merged into a new cluster $C_5$, and all the distances to $C_5$ would be updated in the distance matrix by the corresponding average distances to the sequences in $C_5$. Repeat the same and find new clusters and sequences, merging and updating the distance matrix, until we are left with one cluster.

**Algorithm:** Let the clusters be $C_1$, $C_2$, $C_3$,....., $C_n$ and $s_i$ be the size of each cluster $C_i$, 'd' be the pairwise distance defined on the clusters. Clustering is done in the following manner:

1. Find the smallest pairwise distance amongst the clusters, d ($C_i$, $C_j$).
2. A new cluster $C_k$ with size $s_i + s_j$ is formed by joining $C_i$ and $C_j$.
3. Compute the new distances from all other clusters to $C_k$ by using the existing weighted distances average.

$$d(C_k, C_l) = \frac{(s_i * d(C_i, C_l)) + (s_j * d(C_j, C_l))}{(s_i + s_j)}$$

   Where l € {1, 2... n} and l ≠ i, j.
4. Repeat 1, 2, and 3 until only one cluster remains.

The asymptotic time complexity of UPGMA is O ($n^2$), since there are (n-1) iterations, with O (n) steps per iteration [11].

### 2.2.2. Neighbor-Joining

The Neighbor-Joining algorithm is consistent with a parsimonious evolutionary model in which the total sum of branch lengths is minimized [9]. It has the added benefit of achieving the correct tree topology when given the correct pairwise distances, while also being flexible enough to accommodate many distance models. Now let us look at the actual algorithm.

**Algorithm:** Let the clusters be $C_1$, $C_2$, $C_3$,....., $C_n$ and $s_i$ be the size of each cluster $C_i$, 'd' be the pairwise distance defined on the clusters. Clustering is done in the following manner:

1. Compute

$$u_i = \sum_{k \neq i} \frac{d(C_i, C_k)}{(no.of clusters - 2)}$$

2. Find the smallest difference d $(C_i, C_j)$ – $u_j$ – $u_j$ and choose i and j accordingly.
3. A new cluster $C_k$ is formed by joining $C_i$ and $C_j$. Calculate the intermediate distances between $C_i$, $C_j$ and $C_k$ as:

$$d(C_i, C_k) = \frac{1}{2}[d(C_i, C_j) + u_i - u_j]$$

$$d(C_i, C_k) = \frac{1}{2}[d(C_i, C_j) + u_j - u_i]$$

4. Compute distances between new cluster and all the other cluster's:

$$d(C_k, C_l) = \frac{d(C_i, C_l) + d(C_j, C_l) - d(C_i, C_j)}{2}$$

Where l € {1, 2... n} and l ≠ i, j.
5. Repeat 1, 2 and 3 until you are left with two clusters.

Similarly to UPGMA, the asymptotic time complexity of the Neighbor-Joining algorithm is also O ($n^2$). However, as we shall see, their execution time performance is certainly not identical.

## 2.3. Dynamic Programming

Needleman and Wunsch's elegant algorithm for comparing two protein sequences also works well for aligning nucleic acid sequences. The algorithm actually belongs to a very large class of algorithms for finding optimal solutions called dynamic programming [1]. In dynamic programming a global alignment of two sequences is done based on constructing a scoring matrix. Paths in the scoring matrix decide the optimal solution for two aligned sequences. The scoring matrix is dependent on three variables; match score, mis-match score, and gap penalty. Match indicates that the two letters are the same, mismatch denotes that the two letters are different, and gap (Insertion or Deletion) denotes one letter aligning to a gap in the other string.

Let us consider an example with two strings: ATGCG and TGCAT. Consider the scores as match score=3, mis-match score= 1, and gap penalty= -1. There is no rule that we should have only one optimal solution of aligned sequences. Figure 2 shows the complete scoring matrix obtained after scoring with match, mis-match, and gap scores.

|   |    | T  | G  | C  | A  | T  |
|---|----|----|----|----|----|----|
|   | 0  | -1 | -2 | -3 | -4 | -5 |
| A | -1 | 1  | 0  | -1 | 0  | -1 |
| T | -2 | 2  | 2  | 1  | 0  | 3  |
| G | -3 | 1  | 5  | 4  | 3  | 2  |
| C | -4 | 0  | 4  | 8  | 7  | 6  |
| G | -5 | -1 | 3  | 7  | 9  | 8  |

Figure 2: Dynamic Programming pairwise sequence alignment.

The best optimal alignment that is achieved from above figure are:

A T G C G _
_ T G C A T

## 2.4. Profiles

A profile is a table that records the frequency of each letter at each position in a DNA sequence. Usually profiles are represented using a matrix with letters as columns and position as rows. To build a profile we need at least two DNA sequences. Profiles allows us to identify consensus sequences for two or more sequences that are already aligned. Progressive alignment uses profiles to compute multiple alignment heuristically [14]. In this project, the concept of aligned profiles is used to compute multiple alignments. Below is an example that explains the concept of profiles in detail.

Consider four DNA sequences that are in a MSA: AGT_C, AGTGC, ATTG_ and TG_GT.
The profile for the above four sequences looks like;

|   | 0   | 1   | 2   | 3   | 5   |
|---|-----|-----|-----|-----|-----|
| A | .75 | -   | -   | -   | -   |
| C | -   | -   | -   | -   | .5  |
| G | -   | .75 | -   | .75 | -   |
| T | .25 | .25 | .75 | -   | .25 |
| – | -   | -   | .25 | .25 | .25 |

Figure 3: Profile Scoring Matrix

**Construct a profile sequence:** Depending on the highest score for each position, the character at that position [14] is determined. Suppose we have, at position 0 a 0.75 chance to get `A' and 0.25 chance to get `T'. As the probability of `A' is greater than `T', `A' is chosen at position 0 in the profiled sequence. Moving on the same way `G', `T', `G', and `C' are chosen for the next four positions. Then the final profiled sequence achieved is "AGTGC". This example has no complications in deciding the character, but what if two alphabets have same probability of occurrence? For those situations, we state and implement two simple rules in this work. Consider the character at the same position in the sequence or profile that you want to align to the existing profile. Then, the rules are as follows:

1. If the character exists at that position, compare it with the highest probable characters. If matched, consider the matched character for the profiled sequence; otherwise, select any of the identically highest probable characters with a random draw.
   Example: Consider that `A', `C', `G' and `T' are with 0.25 probability at position 4 in the profile and we have a `-' in the sequence at position 4. Then a random draw is made among `A', `C', `T' and `G'. Suppose that `C' is randomly drawn, then `C' will be at position 4 in the profiled sequence. If any of `A', `C', `G' and `T' are at that position then it will be selected and copied to position 4 of profiled sequence.
2. If the character does not exists at that position, then randomly choose any of the highest probable character.

## 3. EXPERIMENTAL SETUP AND RESULTS

### 3.1. Experimental Setup

We have discussed how to build an effective MSA through the progressive alignment method in section 2. Now we discuss our implementation and results with some sample data sets. For experimental purposes, three types of data sets are used:

**Input:**

1. Seven short sequences each with lengths between 4 and 40. These sequences do not resemble any species but are used for testing.
2. Five sequences with lengths between 40 and 500, also not derived from actual species.
3. The beta-cassein genetic sequences of five mammalian species Rat, Camel, Dog, Whale, and Porpoise.

**System Configuration:** Intel CORE i5 with 4GB RAM and 512GB Hard disk space.

**Operating System:** Ubuntu 14.04

**Programming Language:** C++

**Setup 1:**  **Setup 2:**
Match Score: 3  Match Score: 3
Mis-match Score: 0  Mis-match Score: 0
Gap penalty: -1  Gap penalty: -1
Guide-tree: UPGMA  Guide-tree: Neighbor-Joining

### 3.2. Experimental Results

The Figures 4 and 5 are the guide trees for both UPGMA and Neighbor-Joining, respectively, with input as the large sequences of rat, camel, dog, whale, and porpoise. From these figures, we can clearly observe that the guide tree topologies are not same when different evolutionary tree construction algorithms are used. This has a great impact on the multiple sequence alignment. The progressive multiple alignment is purely dependent on the guide tree topology. Notably, the guide tree topology of Figure 5 for the Neighbor-Joining algorithm is more consistent with the actual phylogenetic tree of these species than that of the UPGMA algorithm.

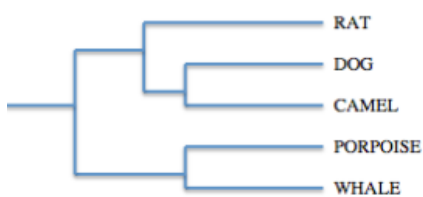 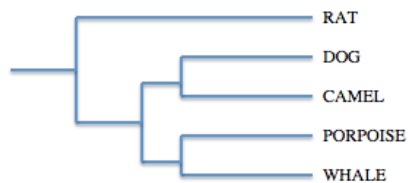

Figure 4: UPGMA　　　　　　　　Figure 5: Neighbor-Joining

Figure 6 shows graphs for execution time generated by considering short and medium length sequences as inputs for UPGMA and Neighbor-joining. Observations from the graph: Both UPGMA and Neighbor-Joining take the same amount of time when the input sequences are of short length. This means that the algorithm used to build the guide tree has no impact on the execution time or time complexity of the program when the progressive alignment is done for relatively shorter sequences.

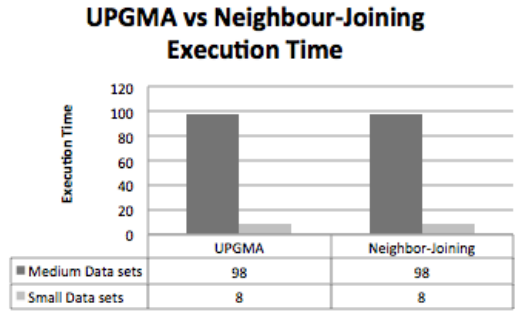

Figure 6: Execution Time for Small sequences

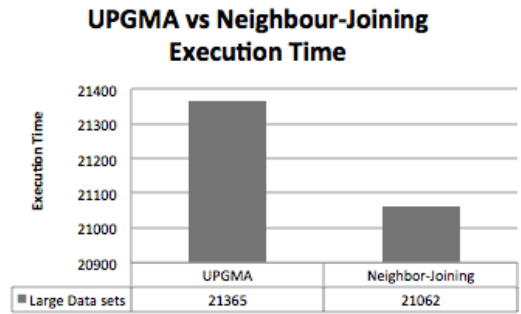

Figure 7: Execution Time for Large sequences

Figure 7 graphs execution times generated by considering large sequences as inputs for UPGMA and Neighbor-joining. Recall that the sample data tested is for genetic sequences of rat, dog, camel, porpoise, and whale. The interesting observation from the graph is that the execution time for both UPGMA and Neighbor-Joining differ. Neighbor-Joining clearly outperforms UPGMA in terms of efficiency in this experiment of a set of long sequences.

Setup 1 with small input sequences:	The resulted sequences are:

ACGTACT	ACGTAC_T_ _ _ _ _ _ _ _
ACTACG	A_CTAC_G_ _ _ _ _ _ _ _
ATGGATACTAACTCGG	ATGGATACTAACTCGG
ATGGCTA_GT	ATGG_ _ _CTA_GT_ _ _
ATGCTCCGGCAAAGG	ATGCTCCGGCAAAGG_
ATGCTGG	ATGCT_ _ _GG_ _ _ _ _ _
ATCGACAGTGTCA	ATCGACAGT_ _GTCA_

Thus Multiple Sequence Alignment is achieved.

Figure 8 below indicates the graph for total alignment costs generated by considering small, medium, and large sequences as inputs for UPGMA and Neighbor-Joining. Observations from the graph are as follows: Neighbor-Joining has lower optimal costs than UPGMA for all types of input sequences. Note that both algorithms attempt to achieve the lowest total cost, meaning the actual optimal cost. Therefore, the results indicate that the final MSA achieved by Neighbor-Joining is more accurate than that achieved by UPGMA.

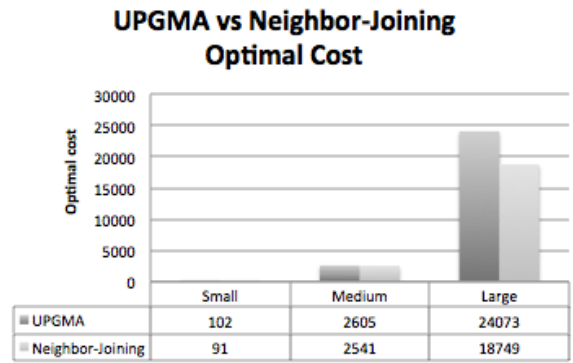

Figure 8: Total Costs for small, medium, and large sequences

## 4. CONCLUSIONS

Progressive alignment is highly dependent on how the guide tree is built and how the sequences are arranged by following the leaf nodes of a guide tree. In this project, we have compared two common guide tree algorithms, UPGMA and Neighbor Joining, in terms of both efficiency and accuracy. A simple and greedy approach is used for profile to sequence and profile to profile alignment, making the progressive alignments relatively faster. Although the five step progressive alignment process previously outlined is not new, some aspects of the profile evaluation and profile alignment used in this project are novel. Furthermore, our experimental results clearly indicate that Neighbor-Joining is superior to UPGMA in both time complexity and final alignment, for use in guide trees for progressive MSA.

## ACKNOWLEDGEMENTS

We would like to take this opportunity to thank Dr. Xudong William Yu and Dr. Mark McKenney for taking their valuable time in reviewing the project and providing good feedback.

**Authors**

**Dega Ravi Kumar Yadav** born in India, in 1991. Perusing Master's in Computer Science at Southern Illinois University, Edwardsville, Illinois, USA. In 2013 published an International Journal titled "Authenticated Mutual Communication between two nodes in MANETs". Areas of interest are Mobile Ad-Hoc networks, Bio-Informatics and AI. 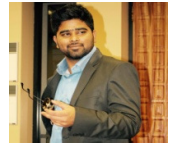

**Dr. GunesErcal,** is an Assistant Professor of Computer Science at Southern Illinois University Edwardsville. Her primary research area is graph theory and its applications to diverse areas such as network analysis and unsupervised learning. She received her PhD in Computer Science from the University of California, Los Angeles in 2008. She received her BS degrees in both Mathematics and Computer Science in 2001 from the University of Southern California. 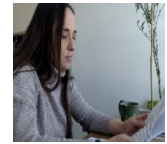